\documentclass[aps,prb,twocolumn,groupedaddress,showpacs,superscriptaddress,amssymb,amsmath,noeprint]{revtex4-2}
\usepackage{amsmath}
\usepackage{float}
\usepackage{graphicx}
\usepackage{dcolumn}
\usepackage{bm}
\usepackage{hyperref}
\usepackage{cleveref}
\usepackage{multirow}
\usepackage{braket}
\hypersetup{
    colorlinks=true,
    linkcolor=blue,
    urlcolor=blue,
    citecolor=blue
}
\usepackage[normalem]{ulem}

\newcommand{\la}{\langle}
\newcommand{\ra}{\rangle}

\newcommand{\mk}[1]{\textcolor{red}{#1}}
\newcommand{\bka}[1]{\textcolor{magenta}{#1}}

\begin{document}

\title{Dynamics of number entropy for  free fermionic systems in presence of defects and stochastic processes}

\author{Atharva Naik}
\email{atharva.naik@icts.res.in} 
\affiliation{International Centre for Theoretical Sciences, Tata Institute of Fundamental Research,
Bangalore 560089, India}

\author{Bijay Kumar Agarwalla}
\email{bijay@iiserpune.ac.in}
\affiliation{Department of Physics, Indian Institute of Science Education and Research Pune, Dr. Homi Bhabha Road, Ward No. 8, NCL Colony, Pashan, Pune, Maharashtra 411008, India}

\author{Manas Kulkarni}
\email{manas.kulkarni@icts.res.in} 
\affiliation{International Centre for Theoretical Sciences, Tata Institute of Fundamental Research,
Bangalore 560089, India}
\thanks{\textsuperscript{†}These authors contributed equally to this work.}

\date{\today} 

\begin{abstract}  
We investigate the dynamics of number entropy in a chain of free fermions subjected to both defects and stochastic processes. For a special class of defects, namely conformal defects, we present analytical and numerical results for the temporal growth of number entropy, the time evolution of the number distribution, and the eigenvalue profile of the associated correlation matrix within a subsystem. We show that the number entropy exhibits logarithmic growth in time, originating from the Gaussian structure of the number distribution. We find that the eigenvalue dynamics reveal a profound connection to the reflection and transmission coefficients of the associated scattering problem for a broad range of defects. When stochastic processes are introduced, specifically Stochastic Unitary Processes (SUP) and Quantum State Diffusion (QSD), the number entropy scales as $\ln(t)$ in the SUP case and shows strong hints of $\ln [\ln(t)]$ scaling in the QSD case. These findings establish compelling evidence that number entropy grows logarithmically slower than the corresponding von Neumann entanglement entropy across a wide class of systems.

\end{abstract}

\maketitle

\textit{Introduction.}--
Understanding the dynamics of various types of quantum correlations in many body quantum systems has gathered tremendous attention over the past several decades, both theoretically~\cite{MCSV2009,MWFCS2010,BBBPM2013,WC2013} and experimentally~\cite{Cheneau2012,Jurcevic2014,Lanyon2017,Görg2018,Rispoli2019}. Such an in-depth understanding is not only important from a fundamental perspective but also from the point of view of making advances in designing quantum materials and technologies~\cite{Keimer2017,ISPS2023,LCDT2025}. One of the widely considered measures of quantum correlation is the entanglement entropy (EE)~\cite{GK2006,KP2006,ECP2010,Jiang2012,LV2021} between a system and its complement which has also been experimentally accessed~\cite{Islam2015,LRSTKCKLG2019,BEJVMLZBR2019,Noel2022,Tajik2023,Koh2023,Lin2024}. It is worth noting that despite experimental progress, the EE is notoriously difficult to measure, especially for extended systems, due to exponential growth in the Hilbert space dimension. Interestingly, however, for 
a certain class of systems, one can infer about entanglement properties by analyzing statistics of conserved charges. More precisely, in the presence of global symmetry (for e.g., total particle number conservation), the state of a subsystem takes a block diagonal form. This enables one to write the EE as a sum of the so called 
number entropy (NE) and configurational entropy (CE)~\cite{LRSTKCKLG2019}. This further makes number entropy a promising candidate to understand 
quantum correlations~\cite{Capizzi_2020,CDM2021,MBC2021,MCP2022,HFSR2023}, phases of matter~\cite{LRSTKCKLG2019,LL2020,KUFMS2020,MRMJ2021,GZ2022, SZ2022,AACKTHBJ2024}, etc. 

NE between a system and its complement captures how particle number fluctuations across a boundary impact the spread of quantum correlations in a system. In other words, it measures the entropy associated with the distribution of particle numbers in a given subsystem and thus serves as a direct probe of number entanglement—the correlations between particle numbers in different regions of the system. Unlike other forms of entanglement that require detailed knowledge of particle configurations, NE focuses solely on how particles are shared between subsystems, making it a robust and accessible diagnostic of quantum dynamics~\cite{GS2018,BRC2019,FG2020,PBC2021,jones2022,FAC2023}. 
Because it isolates number fluctuations from more complex configurational correlations, number entropy helps distinguish regimes where entanglement spreads locally versus globally.  This makes it a powerful measure for understanding fundamental properties of quantum systems, from thermalizing regimes to localized phases, while emphasizing particle number correlations as a primary signature of entanglement.

Despite enormous interest and progress, a thorough analytical and numerical understanding of the dynamics of number entropy is still missing. For example, Ref.~\cite{KUSF2020} provides a bound on NE in terms of EE for non-interacting systems and found that NE is logarithmically smaller both in time and magnitude in comparison to the corresponding EE. Other relevant studies on many body disordered systems~\cite{LL2020,KUFMS2020,MRMJ2021,GZ2022, SZ2022,AACKTHBJ2024} also supported this intriguing connection. However, a rigorous understanding of such a logarithmic smallness in NE so far remains elusive. 

\begin{figure}
    \centering
    \includegraphics[width=\linewidth]{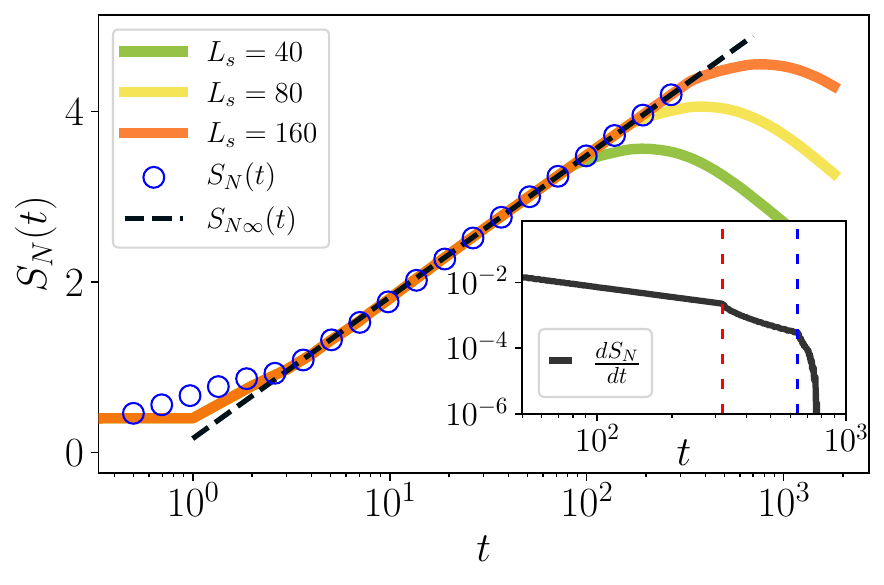}
    \caption{NE [$S_N(t)]$ as a function of time [$t$] for system size $L_s = 40,80$, and $160$ with $g = 0.5$ and $g_c = 0.3$. The solid line shows the exact numerical results. The blue circled line shows the analytically obtained NE given by Eq.~\eqref{eq:S_Nconformal}. The black dashed line is given [Eq.~\eqref{eq:sinf}] by $S_{N\infty}(t) = \frac{1}{2}\log_2 \big[2\pi e\,v\,t \,\lambda^2(1-\lambda^2)\big]$, where $v = 2g/\pi$ and $\lambda = g_c/g$. The analytical result and $S_{N\infty(t)}$ are consistent with the numerical solution up to time $L_s/g$. The inset shows $\frac{dS_N}{dt}$ as a function of time for numerically obtained $S_N(t)$ with $L_s = 160$, which is represented by the solid orange line in the main plot. The vertical dashed lines mark the time when non-analyticity is observed. Red vertical dashed line corresponds to $t = 320$ and blue vertical dashed line corresponds to $t = 640$.}
    \label{fig:Sn_Conformal}
\end{figure}

\begin{figure}
    \centering
    \includegraphics[width=\linewidth]{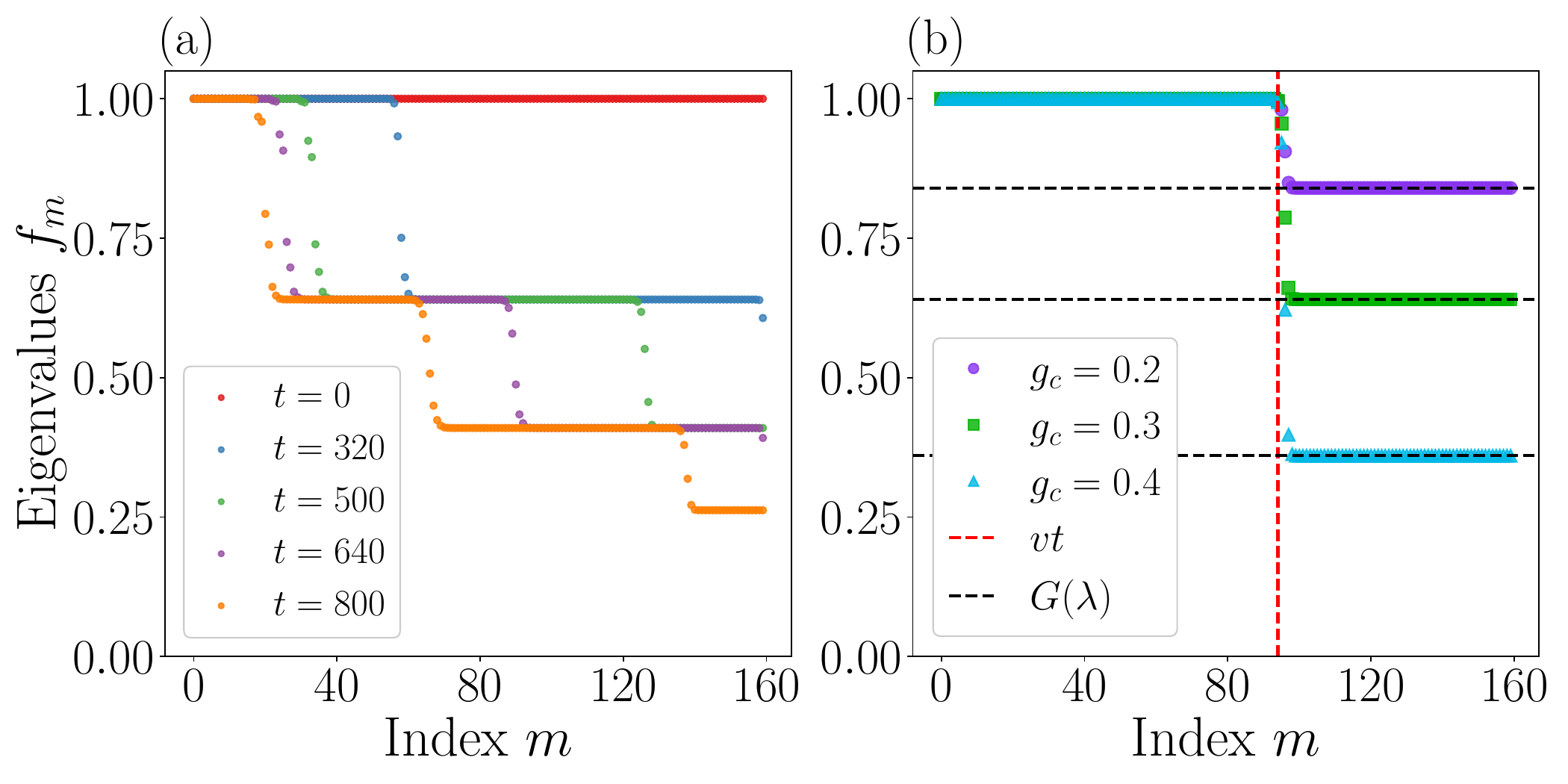}
    \caption{(a) Eigenvalue profile for the system part of the correlation matrix (obtained by ordering the eigenvalues in descending order) at different times for $g_c = 0.3$, $g = 0.5$, $L_s = 160$, $L = 2000$. As the system is quenched, the values of the eigenvalue decrease from 1  and saturate at $(1-\lambda^2)$, where $\lambda = g_c/g$. It can be seen that at times $t = 320$ (blue) and  $t = 640$ (purple), there is a transition in the values. These times correspond to $t = nL_s/g$ for $n \in \mathbb{N}$. The saturation values are given by $(1-\lambda^2)^n$ for $n = 1, 2$ and $3$. (b) Eigenvalue profile for the system part of the correlation matrix for $g_c = 0.2,0.3,0.4$ with $L_s = 160$, $g=0.5$, and at a given time snapshot $t = 200$.
    The black dashed lines mark the saturation value of the eigenvalue profile given by $G(\lambda) = 1-\lambda^2$, where $\lambda = g_c/g$. The vertical red line marks the position of the front, which is given by $(L_s - v\,t)$ where $v = 2g/\pi$ and is independent of $g_c$.}
    \label{fig:eig_prof_conf}
\end{figure}

\begin{figure}
    \centering    \includegraphics[width=\linewidth]{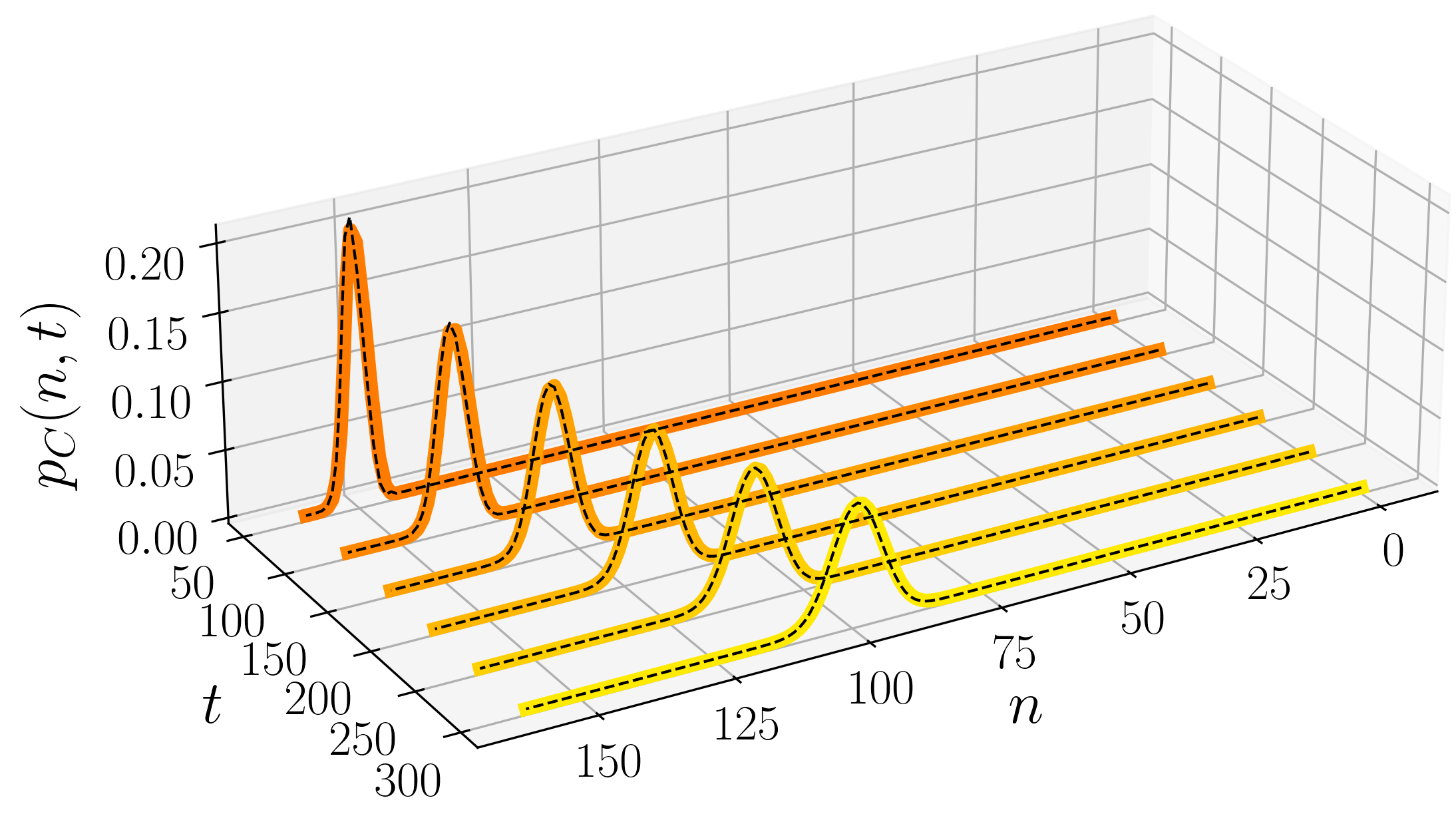}
    \caption{Evolution of the PDF of the particle number $[p_C(n,t)]$ in the system for the setup with a conformal defect, starting with domain wall initial condition. The solid lines are the numerically obtained distribution from the correlation matrix. The black dashed lines are obtained by plotting Eq.~\eqref{eq:p_conform} for the corresponding time values. Here, $L_s = 160$, $g = 0.5$ and $g_c = 0.4$.}
    \label{fig:PDF_Conf}
\end{figure}

\begin{figure}
    \centering
    \includegraphics[width=\linewidth]{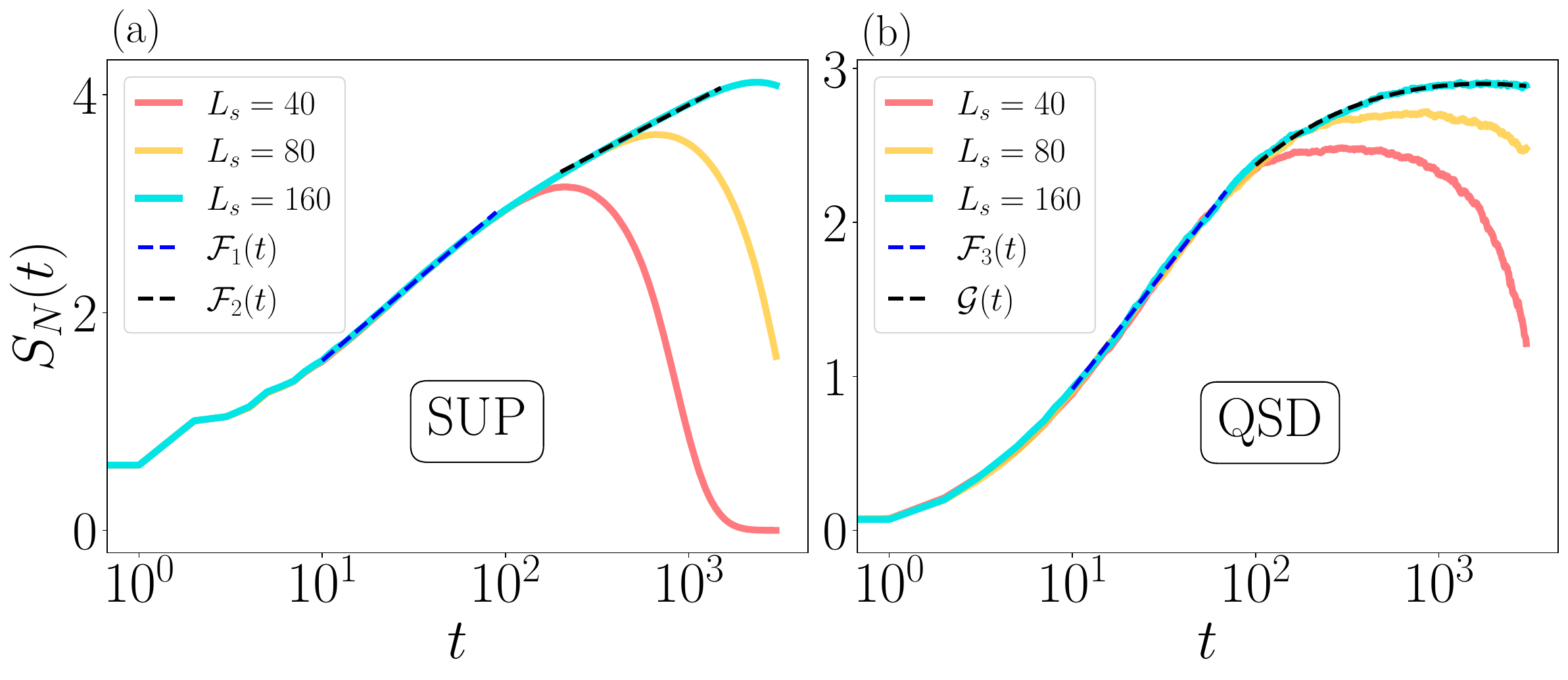}
    \caption{$S_N(t)$ vs $t$ when the probes are present on the system sites for different system sizes. The data is averaged over 100 Quantum Trajectories. The solid lines show the numerically obtained data, the blue dashed line shows the early time scaling, and the black dashed line shows late time scaling. The functions plotted are $\mathcal{F}_n(t) = a_n\ln(t) + b_n$ for $n=1,2,3$ and $\mathcal{G}(t) = a_4\,\ln [\ln(t)] + b_4 \ln(t) + c$. (a) Probes under SUP with $\gamma = 0.1$, $g_c = 0.4$ and $g = 0.5$. Two logarithmic scalings with time, given by $\mathcal{F}_1(t)$ and $\mathcal{F}_2(t)$, are observed. $a_1 = 0.62 $, $b_1 = 0.12$, $a_2 = 0.38$ and $b_2 = 1.26 $. (b) Probe under QSD protocol with $\gamma = 0.05$, $g_c = 0.1$ and $g = 0.5$. At early times, logarithmic growth with time, marked by $\mathcal{F}_3(t)$, is observed. At later times, the growth slows down to a form involving a double logarithmic scaling with time, which is given by $\mathcal{G}(t)$. $a_3 = 0.66$, $b_3 = -0.61$, $a_4 = 5.32$, $b_4 = -0.71$ and $c = -2.46$.}
    \label{fig:Sn_Probe}
\end{figure}

In this work, we first consider a free fermionic lattice setup with a defect and provide numerical and detailed analytical insight about the growth of NE (Fig.~\ref{fig:Sn_Conformal}). We highlight the relation of eigenvalues of the correlation matrix to the transport properties (Fig.~\ref{fig:eig_prof_conf}) and, using it, analytically obtain the evolution of the probability distribution function of the particle number of a subsystem (Fig.~\ref{fig:PDF_Conf}). We further extend the study of NE to cases where the lattice is subjected to different kinds of stochastic processes (Fig.~\ref{fig:Sn_Probe}).  

\textit{Setup.}--We consider a one-dimensional (1D) free fermion chain with the Hamiltonian given as (we set $\hbar=1$ throughout the manuscript),
\begin{equation}
\hat{H} = \sum_{i = -L_s}^{L} g_i(\hat{c}^{\dagger}_i \hat{c}_{i+1} + \textrm{h.c}) + A\sqrt{g^2 - g_c^2}\,(\hat{c}^{\dagger}_0 \hat{c}_{0} - \hat{c}^{\dagger}_{1} \hat{c}_{1}),
\label{eq:Ham}
\end{equation}
where $\hat{c}_i$ and $\hat{c}^{\dagger}_i$ are fermionic annihilation and creation operators, respectively, at site $i$. The hopping constant at site $i$ is given by,
\begin{equation}
    g_i = \begin{cases}
        & -g \hspace{3em} \forall \,\, i\ne 0\\
        & -g_c\hspace{3em} \, i =  0\\
    \end{cases}
\end{equation}
and depending on the value of $A$, different types of defects can be emulated. 
In this work, we mainly focus on two types of defects: (i) conformal defect for which $A=1$, and (ii) hopping defect for which $A=0$. 
We will refer to the first $L_s$ sites as the system and the remaining $L \gg L_s$ sites as the reservoir. The system and the reservoir are always coupled via a defect. In the main text, we will focus on the conformal defect coupling between the system and the reservoir. The results for the non-interacting Hamiltonian with a hopping defect will be discussed in the supplementary material (SM)~\cite{supp}. We choose the domain wall configuration (i.e., filled system and empty reservoir) as our initial state and evolve the setup unitarily under the given Hamiltonian. We will study the dynamics of the eigenvalue profile of the correlation matrix and subsequently investigate the behavior of NE as a function of time.

\textit{Number Entropy .--} Suppose a system has a global particle number conservation symmetry, with total $N$ particles, it can be shown that the reduced density matrix for a sub-system $\rho_s(t)$ takes a block diagonal structure with each block corresponding to a conserved charge~\cite{LRSTKCKLG2019,supp}
\begin{equation}
    \rho_s(t) = \bigoplus_{n=0}^{N}\, p(n,t)\rho^{(n)}_s(t),
    \label{eq:block_diag_rho}
\end{equation}
where $p(n,t)$ is the distribution of the total number of particles in the sub-system at a given time and $\rho^{(n)}_s(t)$ is the block of the reduced density matrix corresponding to the conserved charge $n$. 
In such cases, the von Neumann EE can be written as a sum of NE and CE~\cite{LRSTKCKLG2019,supp}.
\begin{equation}
    S_{vN}(t) = S_N(t) + S_C(t)
\end{equation}
with,
\begin{eqnarray}
    \label{eq:SN}
    S_N(t) &=& - \sum_{n=0}^{N} p(n,t)\log_2\big[p(n,t)\big], \\
    \label{eq:SC}
    S_C(t) &=& - \sum_{n=0}^{N}p(n,t)\sum_i \rho^{(n)}_{ii} \log_2 \big[\rho^{(n)}_{ii}\big],
\end{eqnarray}
where $\rho^{(n)}_{ii}$ are the diagonal elements of the reduced density matrix of the block corresponding to conserved charge $n$ in the Schmidt basis. The Probability Distribution Function (PDF) of the number of particles in the system, $p(n,t)$, can be obtained by taking the inverse Fourier transform of the Characteristic Function (CF), $\chi(\xi,t) = \braket{e^{i\xi \hat{N}}}_t$, where $\hat{N} = \sum_{i=-L_s}^{-1}\hat{c}_i^\dagger\hat{c}_i$ is the total number operator of the system. The CF is related to the eigenvalues, $f_m(t)$, of the system part of the correlation matrix, ${C}_{ij}(t) = \braket{\hat{c}^\dagger_i(t)\hat{c}_j(t)}$ as \cite{PhysRevLett.134.067101,PhysRevLett.110.060602,PhysRevB.75.205329,PhysRevB.103.L041104},
\begin{equation}
    \chi(\xi,t)= \prod_{m=1}^{L_s}\Big{(}1 + (e^{i\xi}-1)f_m(t)\Big{)}.
    \label{eq:chi}
\end{equation}
Thus, knowing $f_m(t)$ enables us to determine the CF [$\chi(\xi,t)$], the PDF [$p(n,t)$], and consequently, the NE [$S_N(t)$]. Calculating CE from the first principles is not as straightforward and is determined by subtracting NE from EE. The EE can be determined using the eigenvalues of the correlation matrix as,
\begin{align}
\label{eq:EE}
        S_{vN}(t) = -\sum_{m=1}^{L_s}&\Big{[}f_m(t)\log_2 \big(f_m(t)\big) \notag\\
        &+ \big(1-f_m(t)\big)\log_2 \big(1-f_m(t)\big)\Big{]}.
\end{align}
Then, using Eq.~\eqref{eq:SN} and Eq.~\eqref{eq:EE}, CE can be obtained  as  $S_C(t) =  S_{vN}(t) - S_N(t)$.

\textit{Evolution of eigenvalues of correlation matrix.--} As discussed in the previous section, we can compute NE by knowing the eigenvalue profile of the system part of the correlation matrix. Recall that we consider a 1D free-fermionic setup with a conformal defect that couples the system and the reservoir. This corresponds to taking $A = 1$ in the Hamiltonian given in Eq.~\eqref{eq:Ham}. It has been shown, in Ref.~\cite{EP2012}, that for a setup with system and reservoir of same size, the correlation matrix of the system part in the presence of a conformal defect is related to the correlation matrix of the system part in the absence of the defect as,
\begin{equation}
    C_{ij}^{C}(t) = \lambda^2 \, C^{H}_{ij}(t) + (1-\lambda^2) \, \delta_{ij},
    \label{eq:Cc}
\end{equation}
where $C_{ij}^{C}(t)$ is the correlation matrix for the setup with a conformal defect, $C_{ij}^{H}(t)$ is the correlation matrix in the absence of the defect, and $\lambda = g_c/g$. The superscript $C$ and $H$, in Eq.~\eqref{eq:Cc}, stand for conformal and homogeneous, respectively. As discussed in Ref.~\cite{EP2012}, diagonalizing the matrices on both sides of the Eq.~\eqref{eq:Cc}, we get,
\begin{equation}
    f_m^{(C)}(t) = \lambda^2 f^{(H)}_{m}(t) + (1-\lambda^2),
    \label{eq:eig_conf}
\end{equation}
where $f^{(C)}_m(t)$ and $f^{(H)}_m(t)$ are the eigenvalues of the system part of the correlation matrix for the setup in the presence and absence of a conformal defect, respectively. Furthermore, interestingly, $(1-\lambda^2)$ and $\lambda^2$ can be identified as the reflection and transmission probabilities, respectively, for a plane wave propagating along the 1D lattice and scattered by a conformal defect, see SM~\cite{supp}. 

The exact analytical form of $f_m^{(H)}(t)$ is not known, and moreover, in our case, the system and the reservoir are not of equal size. Consequently, Eq.~\eqref{eq:eig_conf} cannot be applied directly to obtain $f_m^{(C)}(t)$. Nevertheless, Eq.~\eqref{eq:eig_conf} provides an important insight--- namely, the eigenvalues of the correlation matrix are closely related to the reflection and transmission probabilities. Motivated by this observation, we numerically investigate the time evolution of the eigenvalue profile of the system part of the correlation matrix, and try to infer the functional form of $f^{C}_m(t)$.

The eigenvalue profile of the system part of the correlation matrix for the setup with a conformal defect is shown in Fig.~\ref{fig:eig_prof_conf}. As seen in Fig.~\ref{fig:eig_prof_conf}a, initially, all the eigenvalues are $1$. This is because the initial state we choose to be in a domain wall configuration. When the system-reservoir setup is quenched, there is a flow of particles from the system to the reservoir. As a result of this, an increasing number of eigenvalues take values less than one. As shown in Fig.~\ref{fig:eig_prof_conf}b, there is a transition front of eigenvalues moving from right to left as more eigenvalues are transitioning from the value $1$ and settling down on some constant value. This front moves with a velocity $v = 2g/\pi$. Upon inspection, we can observe that the eigenvalues saturate at the value which is given by $R(m) = 1-\lambda^2$, where $m$ is the index number of the eigenvalues in descending order. Note that $R(m)$ is the same function as the reflection probability for a conformal defect (see SM~\cite{supp}). Further evolving the eigenvalue profile, we see that after time $t_1 = L_s/g$, the eigenvalues at the value $R(m)$ starts transitioning to value $R(m)^2$. Likewise, at time $t_2 = 2L_s/g$, another transition of eigenvalues from $R(m)^2$ to $R(m)^3$ starts, as seen in Fig.~\ref{fig:eig_prof_conf}a. We recall that $R(m)$ is a constant and equal to $(1-\lambda^2)$. Thus, there is a cascading (sharp drop) and saturation of the eigenvalues at the values given by $R(m)^{(q+1)}$ which starts at time $t_q = qL_s/g$. The time $t_1$ corresponds to the minimum time it takes for a particle to go from the defect site, i.e., $i = 0$, to the finite boundary at site $i = -L_s$, reflect, and return back to the defect site again. $q$ denotes the number of reflections off the finite boundary of the system.   
Owing to such a simple nature of the evolution of eigenvalues in the case of the conformal defect, we can analytically determine the number entropy up to a leading order for $t<L_s/g$. We can write the eigenvalue profile, $f^{(C)}_m(t)$, as,
\begin{equation}
    f^{(C)}_m(t) = \begin{cases}
        &1 \hspace{3.4em} 1\le m \le L_s - vt\\
        &1-\lambda^2 \hspace{1em} L_s - vt<m<L_s
    \end{cases}
    \label{eq:f(k,t)}
\end{equation}
where $v = 2g/\pi$ and $\lambda = g_c/g$. Note that Eq.~\eqref{eq:f(k,t)} is a leading order approximation of Eq.~\eqref{eq:eig_conf}. There are still $\mathcal{O}(1)$ number of eigenvalues between the value 1 and $(1-\lambda^2)$ that are not accounted for in Eq.~\eqref{eq:f(k,t)}. These $\mathcal{O}(1)$ number of eigenvalues will contribute to NE only at a sub-leading order and are thus ignored in this analysis. Contribution of these $\mathcal{O}(1)$ number of eigenvalues is apparent in the absence of defect, as shown in SM~\cite{supp}. We emphasize that Eq.~\eqref{eq:f(k,t)} is true only for $t<L_s/g$. Now substituting Eq.~\eqref{eq:f(k,t)} in Eq.~\eqref{eq:chi}, we can get the CF for conformal defect $\chi_C(\xi,t)$ as
\begin{eqnarray}
        \chi_{C}(\xi,t) &=& \prod_{m=1}^{L_s} \Big{(}1 + (e^{i\xi}-1)f^{(C)}_m(t)\Big{)}, \notag\\
        &=& e^{i\xi(L_s - vt)}\Big{(}1 +(e^{i\xi} - 1)(1-\lambda^2)\Big{)}^{vt}.
\end{eqnarray}
Rewriting $\chi_C(\xi,t)$ as,
\begin{equation}
    \label{eq:chiC}
    \chi_C(\xi,t) = e^{i\xi(L_s - vt)}\Big{(}\lambda^2 + (1-\lambda^2)e^{i\xi}\Big{)}^{vt},
\end{equation}
we notice that the factor in the parenthesis in Eq.~\eqref{eq:chiC} is of the form $\chi(\xi,t)  = (q + p \, e^{i\xi})^n$ which is simply the CF corresponding to a binomial distribution $\tilde{p}(k) = \binom{n}{k} \,\,p^k q^{n-k}$ with $q = \lambda^2$, $p = 1-\lambda^2$, and $n = vt$. Taking an inverse Fourier transform of Eq.~\eqref{eq:chiC}, we obtain,
\begin{align}
        p_C(n,t) &= \int_{-\pi}^{\pi}\frac{d\xi}{2\pi}e^{-i\xi n} e^{i\xi(L_s - vt)}\Big{(}\lambda^2 + (1-\lambda^2)e^{i\xi}\Big{)}^{vt},\nonumber \\
        &\!\!\!\!= \int_{-\pi}^{\pi}\frac{d\xi}{2\pi}e^{-i\xi (n-L_s+vt)} \Big{(}\lambda^2 + (1-\lambda^2)e^{i\xi}\Big{)}^{vt}.
    \label{eq:p_int}
\end{align}
Eq.~\eqref{eq:p_int} can be identified as a shifted binomial distribution, which is given by,
\begin{equation}
\!\!\!p_C(n,t) \!=\! \binom{vt}{vt - L_s + n} (\lambda^2)^{(L_s - n)}(1-\lambda^2)^{vt - (L_s - n)}.
    \label{eq:p_conform}
\end{equation}
Fig.~\ref{fig:PDF_Conf} shows the consistency between the the numerically obtained $p_C(n,t)$ and Eq.~\eqref{eq:p_conform}. To obtain the $p_C(n,t)$ numerically, we do the following: we numerically evolve the correlation matrix $C_{ij}$ using the Hamiltonian given in Eq.~\eqref{eq:Ham}. Then, at each time step, we diagonalize the correlation matrix to obtain the eigenvalues of the system part, i.e., $f^{(C)}_m(t)$. We use these numerically obtained $f^{(C)}_m(t)$ in Eq.~\eqref{eq:chi} to get the CF, $\chi_c(\xi,t)$. We then perform an inverse Fourier Transform of CF to get the PDF, $p_C(n,t)$. 

We can use Eq.~\eqref{eq:p_conform} in Eq.~\eqref{eq:SN} to determine the NE as,
\begin{align}
        S_N(t) &= -vt\,(1-\lambda^2)\log_2(1-\lambda^2) - vt\,\lambda^2 \log_2(\lambda^2) \notag\\
        &- \sum_{x = 0}^{vt} \binom{vt}{x}(\lambda^2)^x (1-\lambda^2)^{vt-x}\times\notag\\
        &\Big{[}\log_2((vt)!) - \log_2((vt-x)!) - \log_2(x!)\Big{]}.
    \label{eq:S_Nconformal}
\end{align}
Here, $y! = \Gamma(y+1)$. As mentioned before, this result is only valid for $t<L_s/g$. It can be seen from Fig.~\ref{fig:Sn_Conformal} that Eq.~\eqref{eq:S_Nconformal} is consistent with the numerically obtained $S_N(t)$. Now, if we first take the limit $L_s\xrightarrow{}\infty$ and then take $t\xrightarrow{}\infty$, in Eq.~\eqref{eq:p_conform} we have a binomial distribution 
of the form, $\tilde{p}(k) = {}^nC_k \,\,p^k q^{n-k}$, 
with $n\xrightarrow{}\infty$. It is well known, using the central limit theorem, that the binomial distribution can be approximated as a Gaussian distribution in this limit. Therefore, the probability distribution function $p_C(n,t)$ can now be approximated as,
\begin{equation}
    p_\infty(n,t) = \frac{1}{\sqrt{2\pi vt\lambda^2(1-\lambda^2)}}\exp\bigg[-\frac{\big[n-vt(1-\lambda^2)\big]^2}{2vt\lambda^2(1-\lambda^2)}\bigg].
\end{equation}
Thus, in the thermodynamic limit at late times, the NE takes the form,
\begin{equation}
    S_{N\infty}(t) = \frac{1}{2}\log_2 \big[2\pi e vt \lambda^2(1-\lambda^2)\big] \sim \log_2(t).
    \label{eq:sinf}
\end{equation}
For a finite system, determining the NE analytically for times $t>L_s/g$ is not as straightforward. Even though we know the values at which the eigenvalue profile saturates and the exact time at which the cascading of the eigenvalues is observed, as shown in Fig.~\ref{fig:eig_prof_conf}a, we do not know the velocity at which the different fronts move. Moreover, the form of the CF is not simple enough to determine the PDF. Thus, we need to resort to numerical means to obtain the complete dynamics of NE. The numerical results are shown in Fig.~\ref{fig:Sn_Conformal}. Initially, the $S_N(t)$ is zero as all the particles are in the system. As more particles enter the reservoir, the $S_N(t)$ increases, attains a maximum, and then shows a decay. This kind of behavior of an entropy dynamics is called a Page-curve, and it has been studied extensively in recent years for non-interacting as well as interacting systems in the context of EE~\cite{Tokusumi_2018,Liu2021,Kehrein2024,Glatthard2024,SKD2024,Oishikawa2025,RDK2025,JMK2025,LKG2025,GGNAK2025}. It can be seen from Fig.~\ref{fig:Sn_Conformal} that even for $L_s = 160$, $S_N(t)$ grows as $\sim \ln(t)$, as predicted by Eq.~\eqref{eq:sinf}. This scaling for the growth of NE is logarithmically slower than EE for the same setup, which has been studied numerically and analytically using hydrodynamic techniques in Ref.~\cite{SKD2024}. 

Another interesting feature we observe is that as the finite-size effect of the system kicks in, there is a non-analyticity seen along with a change in the scaling of $S_N(t)$. This non-analyticity is captured by the inset of Fig.~\ref{fig:Sn_Conformal}. The inset shows the rate of change of $S_N$ as a function of time. We observe that the rate of the number entropy generation shows kinks at times which coincide with the cascading events seen in Fig.~\ref{fig:eig_prof_conf}a, i.e., specifically at times $t = 320, 640$. While we have highlighted two of these non-analyticities in our results, additional kinks are expected at times $t = qL_s/g$, with $q > 2$, which can be more pronounced in appropriate parameter regimes. 

\vspace{1em}
\noindent \textit{NE in presence of stochastic processes--} We will next discuss the temporal evolution of NE when the system sites are subjected to stochastic probes. To calculate the NE, we need to carefully unravel the density matrix. To do so, we will use two Quantum Trajectory protocols: (i) Stochastic Unitary Process (SUP) and (ii) Quantum State Diffusion (QSD), which are discussed in detail in Ref.~\cite{GGNAK2025}. First, let's consider the SUP protocol. The SUP protocol corresponds to onsite fluctuating noise added to the Hamiltonian, 
which makes the evolution of a pure state stochastic but unitary ~\cite{CBD2017,DMSD2020,GGNAK2025}. It can be shown that the evolution of a pure state is dictated by a Stochastic Schr\"odinger Equation (SSE) of the form,
\begin{equation}
    d\,\big|\psi_{\xi}(t)\big\rangle \!= \! \big[\!-\!i\hat{H}dt-i\sum_{j=-L_s}^{j = 0}d\xi_j^t\,\hat{n}_j\!-\!\frac{\gamma}{2}\sum_{j=-L_s}^{j =0}\,\hat{n}_j\,dt\,\big]\big|\psi_{\xi}(t)\big \rangle.\label{SSE-SUU1}
\end{equation}
where $d\xi^t_j$ is a Weiner increment with $\bar{d\xi^{t}_i} = 0$ and $d\xi^{t}_i\xi^{t}_j = \gamma dt\delta_{ij}$ and $\gamma$ is the noise strength. The QSD protocol corresponds to making a weak measurement of the local particle number at each site with a measurement rate $\gamma$~\cite{CM1987,DGHP1995,GP1997,Brun2002,Diehl2021,MZ2022,GGNAK2025}. The evolution is governed by the SSE of the form,
\begin{equation}
     d\,\big|\psi_{\xi}(t)\big\rangle\!=\!\! \big[\!-\!i\hat{H}dt+\sum_{j=-L_s}^{j =0}d\xi_j^t\,\hat{M}_j\!-\!\frac{\gamma}{2}\sum_{j=-L_s}^{j =0}\,\hat{M}_j^2 dt\,\big]\,\big|\psi_{\xi}(t)\big \rangle.\label{SSE-QSD}
\end{equation}
where $\hat{M}_j= \hat{n}_j-\la \hat{n}_j\ra_t$ and $\langle \hat{n}_j\ra_t$ is the average computed over the quantum trajectory $|\psi_{\xi}(t)\big\rangle$. We start from a domain-wall initial state and numerically evolve it according to the specified protocols, computing the correlation matrix at each time step. We then diagonalize the system part of the correlation matrix to obtain its eigenvalues and follow the procedures outlined in the preceding sections to compute the NE for a given Quantum Trajectory ($S_N^\xi(t)$). But note that, since the Quantum Trajectories are the stochastic solution of the SSE, the quantities computed using them are also stochastic. Hence, to define a meaningful notion of NE for the stochastic settings, we carry out an ensemble average of $S_N^{\xi}(t)$ over all the trajectories. We name ensemble averaged NE $S_N(t) = \langle S_N^{\xi}(t)\rangle_{\xi}$.

Fig.~\ref{fig:Sn_Probe} shows the time evolution of NE for SUP (Fig.~\ref{fig:Sn_Probe}a) and QSD (Fig.~\ref{fig:Sn_Probe}b) protocols. It was shown in Ref.~\cite{GGNAK2025} that when the stochastic probes are present at the system sites, the EE initially grows linearly in time. Beyond a timescale of order $1/\gamma$, EE grows as $\sqrt{t}$ for the SUP case and as $\ln(t)$ for the QSD case. As shown in Fig.~\ref{fig:Sn_Probe}a, for the SUP case, the NE initial grows as $\ln(t)$, before transitioning to a slower $\ln(t)$ scaling, albeit with different coefficients. Note that the coefficient of the late-time $\ln(t)$ scaling, denoted by $a_2$, is approximately half that of the early-time $\ln(t)$ scaling. This behavior reflects the underlying dynamics of the EE, which grows linearly at early times and crosses over to a $\sqrt{t}$ growth at later times. Contrary to SUP, for QSD, as shown in Fig.~\ref{fig:Sn_Probe}b, NE initially grows as $\ln(t)$ followed by a slower than $\ln(t)$ growth. While we expect the NE to eventually scale as $\ln(\ln(t))$, we were unable to access the regime where this behavior becomes fully manifest. Nevertheless, we see a scaling of the form $a_4\ln(\ln(t)) + b_4\ln(t) + c$, with $a_4, b_4, c \ne 0$, which may signal an emergence of a crossover to the slower $\ln(\ln(t))$ scaling in an appropriate parameter regime.

\textit{Summary.--} In this work, we investigated the growth of number entropy in a fermionic lattice system in the presence of defects and two different kinds of stochastic processes. For the free fermionic case, we established that the eigenvalues of the correlation matrix take the values that correspond to the reflection probability of a plane wave scattered by the given defect. In particular, for a conformal defect, the eigenvalue profile assumes a remarkably simple form, which can be conjectured directly from numerical simulation. This enabled us to analytically show that the probability distribution function of the number of particles in the system is a shifted binomial distribution given by Eq.~\eqref{eq:p_conform}, in excellent agreement with numerical results. Consequently, we were able to show analytically how the logarithmically slower growth of NE emerges in this setting. Furthermore, we numerically obtained the temporal evolution of NE in the presence of two kinds of stochastic processes, namely, SUP and QSD. For all cases, non-interacting as well as stochastic, we find evidence that NE grows logarithmically slower than the corresponding von Neumann EE.

Having established a direct connection between the eigenvalues of the correlation matrix and the reflection/transmission probabilities, an important open question is to understand rigorously how this correspondence arises. A deeper understanding of this relation could enable us to characterize the complete time evolution of the eigenvalue spectrum. It would be particularly interesting to explore whether a universal relation of this kind holds for arbitrary point defects. A natural extension is to investigate whether such a framework can be used for stochastic processes. 

\textit{Acknowledgments.--} We thank Katha Ganguly for the initial progress related to the project.  M.K. and A.N. acknowledge support from the Department of Atomic Energy, Government of India, under project No.~RTI4001. B.K.A acknowledges CRG Grant No.~CRG/2023/003377 from the Science and Engineering Research Board (SERB), Government of India.  
B.K.A thanks the hospitality of the International Centre of Theoretical Sciences (ICTS), Bangalore, India under the associateship program.

\bibliography{references}

\setcounter{equation}{0}
\setcounter{figure}{0}
\renewcommand{\theequation}{S\arabic{equation}}
\renewcommand{\thefigure}{S\arabic{figure}}

\begin{center}
{\textbf{\underline{Supplemental Material}}}
\end{center}

\section{Derivation of number entropy}
\label{sec:sup-NE}

In this section, we discuss the derivation of the Number Entropy (NE) and Configuration Entropy (CE) for a system with global symmetry~\cite{LRSTKCKLG2019}, such as the one described by the Hamiltonian in Eq.~\eqref{eq:Ham} of the main text. We will consider a one-dimensional (1D) free fermionic lattice of size $L$ and partition it such that the first $L_s$ sites are called the ``system" and the remaining $L - L_s$ sites are called the ``reservoir". Note in our case $L\gg L_s$. If the entire setup has a global symmetry, then the reduced density matrix of the system, $\rho_S$, decomposes into a block diagonal structure. We will use this form of the reduced density matrix to write the von Neumann Entanglement Entropy (EE) as a sum of NE and CE. 

In our work, we consider $N$ fermions on a 1D lattice and the specific global symmetry is the conservation of the total number of fermions. We choose $N = L_s$. An arbitrary state, $\ket{\psi(t)}$, can be decomposed as, 
\begin{eqnarray}
\label{supp_eq:psi}
    \ket{\psi(t)} = \sum_{n=0}^N c_n(t) \ket{\phi_n},
\end{eqnarray}
where $\ket{\phi_n}$ is a state in occupation number basis with $n$ fermions in the system and $(N-n)$ fermions in the reservoir. The subspace, $\mathcal{H}^{(n)}$, of the Hilbert space with $n$ particles in the system, which contains the state $\ket{\phi_n}$, decomposes into 
\begin{equation}
    \label{supp_eq:hilbert}
    \mathcal{H}^{(n)} = \mathcal{H}_S^{(n)} \otimes \mathcal{H}_R^{(N-n)},
\end{equation}
where $\mathcal{H}_S^{(n)}$ and $\mathcal{H}_R^{(N-n)}$ are the subspaces of the Hilbert space with $n$ particles in the system and $(N-n)$ particles in the reservoir, respectively. This allows us to write,
\begin{equation}
    \label{supp_eq:decom_phi_n}
    \ket{\phi_n} = \ket{n}\!_S \otimes \ket{N-n}\!_R,
\end{equation}
where $\ket{n}_S$ and $\ket{N-n}_R$ are states with $n$ and $(N-n)$ fermions in the system and reservoir respectively. It is easy to see that ${\rm dim}(\mathcal{H}_S^{(n)}) = \binom{N}{n}\equiv d_S^n$ and ${\rm dim}(\mathcal{H}_R^{(n)}) = \binom{N}{N-n} \equiv d_R^n$. We can expand $\ket{n}_{S}$ as a superposition of different configurations of $n$ fermions in the system as,
\begin{eqnarray}
\label{supp_eq:decom_nS}
        \ket{n}_S &=& \sum_{i=1}^{d_S^n} a^{(n)}_i\ket{i_n}\!_S,
\end{eqnarray}
where $\ket{i_n}_{S}$ is some $i$-th configuration of system with $n$ fermions. In Eq.~\eqref{supp_eq:decom_nS}, the subscript $S$ denotes the system. Similarly, we can expand the state $\ket{N-n}_R$ as a superposition of the different configuration of $(N-n)$ fermions in the reservoir as,
\begin{eqnarray}
\label{supp_eq:decom_nR}
    \ket{N-n}_R &=& \sum_{i=1}^{d_R^n} b^{(n)}_i\ket{i_{N-n}}\!_R,
\end{eqnarray}
where $\ket{i_{N-n}}_R$ is the $i-$th configuration of the reservoir with $(N-n)$ fermions. In Eq.~\eqref{supp_eq:decom_nR}, the subscript $R$ denotes the reservoir. Using Eqs.~\eqref{supp_eq:decom_phi_n}-\eqref{supp_eq:decom_nR}, the state $\ket{\psi(t)}$, given in Eq.~\eqref{supp_eq:psi}, can be written as,
\begin{equation}
    \ket{\psi(t)} =\sum_{n=0}^N c_n(t) \sum_{i=1}^{d_S^n}\sum_{j=1}^{d_R^n} a^{(n)}_ib^{(n)}_j \ket{i_n}_S \otimes \ket{j_{N-n}}_R.
    \label{supp_eq:psi_cn}
\end{equation}
Therefore, using Eq.~\eqref{supp_eq:psi_cn}, the corresponding density matrix at time $t$ is given by,
\begin{eqnarray}
\label{supp_eq:rho}
    \rho(t) &=& \ket{\psi(t)}\bra{\psi(t)} \nonumber \\
        &=& \sum_{n,m=0}^N \!\!\!\!c_n(t) c_m^*(t) \sum_{i=1}^{d^n_S}\sum_{j=1}^{d^n_R}\sum_{k=1}^{d^m_S}\sum_{l=1}^{d^m_R} a^{(n)}_i b^{(n)}_j a^{(m)*}_k b^{(m)*}_l \nonumber \\
        &&\hspace{2.5em}\ket{i_n}\!_S \otimes\ket{j_{N-n}}\!_R\hspace{0.5em} {}_S\!\bra{k_m}\otimes {}_R\!\bra{l_{N-m}}
\end{eqnarray}
Each configuration state in the occupation number basis is orthonormal to each other. In other words,
\begin{align}
\label{supp_eq:orthonormality}
        &{}_S\!\braket{i_n|k_m}_S = \delta_{nm}\delta_{ik}.\\ 
&{}_R\!\braket{j_n|l_m}_R = \delta_{nm}\delta_{jl}.
\label{supp_eq:orthonormalityB}
\end{align}
The reduced density matrix for the system can be obtained by taking a partial trace of the density matrix $\rho(t)$, given in Eq.~\eqref{supp_eq:rho}, over the reservoir.
\begin{equation}
\!\!\rho_s(t) \!=\! \textrm{Tr}_R [\rho(t)] \!=\! \sum_{p=0}^N {}_R\!\braket{N\!-\!p|\psi(t)}\braket{\psi(t)|N\!-\!p}_R
\end{equation}
Using Eq.~\eqref{supp_eq:decom_nR} for $\ket{N-p}_R$, we get,
\begin{align}
    \rho_s(t) = \sum_{p=0}^N\sum_{q=1}^{d^p_R}\sum_{r=1}^{d^p_R}b^{(p)}_r b^{(p)*}_q {}_R\!\braket{q_{N-p}|\psi(t)}\braket{\psi(t)|r_{N-p}}_R.
    \label{supp_eq:part_trace}
\end{align}
Substituting $\ket{\psi(t)}$ from Eq.~\eqref{supp_eq:psi_cn} in Eq.~\eqref{supp_eq:part_trace}, the reduced density matrix becomes,
\begin{align}
    &\rho_s(t) = \sum_{p=0}^N\sum_{q=1}^{d^p_R}\sum_{r=1}^{d^p_R}b^{(p)}_rb^{(p)*}_q \sum_{n=0}^N\sum_{m=0}^N c_n(t) c_m^* (t)\times \notag\\
        &\hspace{3em}\sum_{i=1}^{d^n_S}\sum_{j=1}^{d^n_R}\sum_{k=1}^{d^m_S}\sum_{l=1}^{d^m_R} a^{(n)}_i b^{(n)}_j a^{(m)*}_k b^{(m)*}_l\times \notag\\
        &\ket{i_n}\!_S \otimes {}_R\!\braket{q_{N-p}|j_{N-n}}\!_R\, {}_S\!\bra{k_m} \otimes {}_R\!\braket{l_{N-m}|r_{N-p}}\!_{R}.
\label{supp_eq:expand_rho_S}
\end{align}
Using the orthogonality relations given in Eq.~\eqref{supp_eq:orthonormality} and Eq.~\eqref{supp_eq:orthonormalityB} in Eq.~\eqref{supp_eq:expand_rho_S} will give us a factor of $\delta_{n,p}\delta_{m,p}\delta_{q,j}\delta_{l,r}$, which will reduce the 9 summations in Eq.~\eqref{supp_eq:expand_rho_S} into 5 summations and simplifies the equation as,
\begin{align}
        \rho_s(t)&=\sum_{n=0}^N |c_n(t)|^2 \sum_{i,k=1}^{d^n_S}\sum_{j,l=1}^{d^n_R} |b^{(n)}_j|^2|b^{(n)}_l|^2 \times \notag\\
        &\hspace{2em}a^{(n)}_i a^{(n)*}_k\ket{i_n}\!_S {}_S\!\bra{k_n}.
    \label{supp_eq:rho_s_last_step}
\end{align}
We can concisely rewrite Eq.~\eqref{supp_eq:rho_s_last_step} for the reduced density matrix for the system as,
\begin{equation}
\label{supp_eq:rho_sys}
        \rho_{s}(t) = \sum_{n=0}^N |c_n(t)|^2 B^{(n)} \sum_{i,k=1}^{d^n_S}A^{(n)}_{ik}\ket{i_n}\!_S {}_S\bra{k_n},
\end{equation}
where, 
\begin{equation}
\label{supp_eq:Bn}
    B^{(n)} = \sum_{j,l=1}^{d^n_R} |b^{(n)}_j|^2|b^{(n)}_l|^2 = \Big{(}\sum_{l=1}^{d^n_R} |b^{(n)}_l|^2\Big{)}^2,
\end{equation}
and 
\begin{equation}
\label{supp_eq:An}
    A^{(n)}_{ik} = a^{(n)}_ia^{(n)*}_k.
\end{equation}
By defining,
\begin{align}
    p(n,t) \equiv |c_n(t)|^2
\end{align}
and 
\begin{align}
    \rho_s^{(n)} \equiv B^{(n)} \sum_{i,k=1}^{d^n_S} A^{(n)}_{ik}\ket{i_n}\!_S {}_S\bra{k_n},
\end{align}
 we can rewrite Eq.~\eqref{supp_eq:rho_sys} as
\begin{equation}
    \rho_s(t) = \bigoplus_{n=0}^N p(n,t) \rho_s^{(n)}.
    \label{supp_eq: decomp_rho}
\end{equation}
where $\oplus$ denotes the direct sum. 
Additionally, note that states $\ket{\psi}$ and $\ket{\phi_n}$ in Eq.~\eqref{supp_eq:psi} and Eq.~\eqref{supp_eq:decom_phi_n}, respectively, have two normalization conditions,
\begin{eqnarray}
 \label{supp_eq:norm_psi}\braket{\psi|\psi} &=& 1\\
 \label{supp_eq:norm_phi}\braket{\phi_n|\phi_n} &=& 1.
\end{eqnarray} 
Using the first normalization condition given in Eq.~\eqref{supp_eq:norm_psi} with Eq.~\eqref{supp_eq:psi_cn} we get,
\begin{equation}
\label{supp_eq:norm1}
    \sum_{n=0}^N |c_n(t)|^2 \sum_{i=1}^{d^n_S}\sum_{j=1}^{d^n_R} |a_i^{(n)}|^2 |b_j^{(n)}|^2 = 1
\end{equation}
Using the second normalization condition given in Eq.~\eqref{supp_eq:norm_phi} with Eqs.~\eqref{supp_eq:decom_phi_n}-\eqref{supp_eq:decom_nR}, we obtain,
\begin{equation}
\label{supp_eq:norm2}
    \sum_{i,j=1}^{d^n_R} |a_i^{(n)}|^2 |b_j^{(n)}|^2 = 1
\end{equation}
Substituting Eq.~\eqref{supp_eq:norm2} in Eq.~\eqref{supp_eq:norm1} gives us,
\begin{equation}
    \sum_{n=0}^N |c_n(t)|^2 = 1 = \sum_{n=0}^N p(n,t).    
\end{equation}
Hence, as expected, we can interpret $p(n,t)$ as a probability of having $n$ particles in the system at a given time $t$. We will use the result in Eq.~\eqref{supp_eq: decomp_rho} to obtain the equation for NE and CE. We know that the EE is given by,
\begin{equation}
\label{supp_eq:SvN}
    S_{vN}(t) = -\textrm{Tr}_S\Big{(}\rho_s(t) \log_2(\rho_s(t)\Big{)}.
\end{equation}
Henceforth, we will work in the Schmidt basis. For a bi-partitioned system with a state $\ket{\psi}\in \mathcal{H}_S\otimes\mathcal{H}_R$, there exist orthonormal basis, $\{\ket{u_i}\}\in \mathcal{H}_S$ and $\{\ket{v_i}\}\in \mathcal{H}_R$, such that 
\begin{equation}
\ket{\psi} = \sum_{i=1}^{d} \sqrt{\lambda_i }\ket{u_i}\otimes\ket{v_i},    
\end{equation}
 with $\lambda_i\ge 0$ and $d = \textrm{min}\big[\mathrm{dim}(\mathcal{H}_S),\textrm{dim}(\mathcal{H}_R)\big]$. Basis states $\{\ket{u_i}\}$ and $\{\ket{v_i}\}$ are called Schmidt basis and $\{\lambda_i\}$ are called the Schmidt coefficients which satisfy $\sum_i \lambda_i = 1$. It is easy to see that in this choice of basis, the reduced density matrix for both partitions is diagonal,
\begin{align}
 &\rho_s = \textrm{Tr}_R(\rho) = \sum_{j=1}^d \lambda_j \ket{u_j}\bra{u_j}\\   
 &\rho_R = \textrm{Tr}_S(\rho) = \sum_{j=1}^d \lambda_j \ket{v_j}\bra{v_j}.
\end{align}

In the Schmidt basis, Eq.~\eqref{supp_eq:SvN} takes the form,
\begin{equation}
    S_{vN}(t) = -\sum_{i} \rho_{ii}(t) \log_2(\rho_{ii}(t)),
\end{equation}
where $\rho_{ii}$ are the diagonal elements of the reduced density matrix of the system in the Schmidt basis. Substituting $\rho_s(t)$ from Eq.~\eqref{supp_eq: decomp_rho} in Eq.~\eqref{supp_eq:SvN} and writing it in the Schmidt basis we get,
\begin{align}
        S_{vN}(t) &= -\sum_{n=0}^N\sum_{i} p(n,t)\rho^{(n)}_{ii} \log_2(p(n,t)\rho^{(n)}_{ii})\notag\\
        &= -\sum_{n=0}^N p(n,t)\log_2(p(n,t))\sum_{i}\rho^{(n)}_{ii} \notag\\
        &-\sum_{n=0}^N p(n,t)\sum_i \rho^{(n)}_{ii} \log_2(\rho^{(n)}_{ii}).
\end{align}
Finally, using $\textrm{Tr}(\rho^{(n)}) = \sum_i \rho^{(n)}_{ii} = 1$ we get,
\begin{align}
        S_{vN}(t) &= - \sum_{n=0}^N p(n,t)\log_2(p(n,t)) \notag\\
        &- \sum_{n=0}^Np(n,t)\sum_i \rho^{(n)}_{ii} \ln(\rho^{(n)}_{ii}) \notag\\
        &\equiv S_N + S_C,
\end{align}
where we have defined the number entropy as,
\begin{equation}
    \label{supp_eq:SN}
    S_N(t) = - \sum_{n=0}^N p(n,t)\log_2 \big[p(n,t)\big]
\end{equation}
and the configurational entropy as 
\begin{equation}
    \label{supp_eq:SC}
    S_C(t) = - \sum_{n=0}^Np(n,t)\sum_i \rho^{(n)}_{ii} \log_2 \big[\rho^{(n)}_{ii}\big].
\end{equation}
Eq.~\eqref{supp_eq:SN} and Eq.~\eqref{supp_eq:SC} are required equations for calculating NE and CE, respectively for an arbitrary number conserving system.

\section{Results for Number Entropy for free fermion setup without defect}
\label{sec:sup-No_defect}

\begin{figure}
    \centering
    \includegraphics[width=0.9\linewidth]{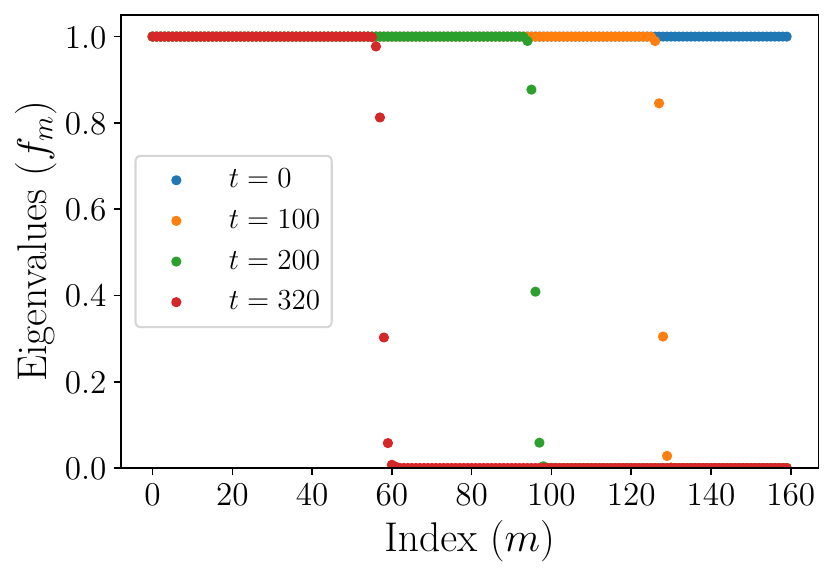}
    \caption{Evolution of the eigenvalue profile of the system part of the correlation matrix for a homogeneous setup with no defect [Eq.~\eqref{supp_eq:Ham} with $g = g_c$]. Here, we have chosen $L_s = 160$ and $g = 0.5$. As clearly seen in the figure, among the $L_s$ eigenvalues, only $\mathcal{O}(1)$ of them are different from 1 and 0, which contribute to entanglement.}
    \label{supp_fig:no_defect_eig_prof}
\end{figure}
\begin{figure}
    \centering
    \includegraphics[width=0.85\linewidth]{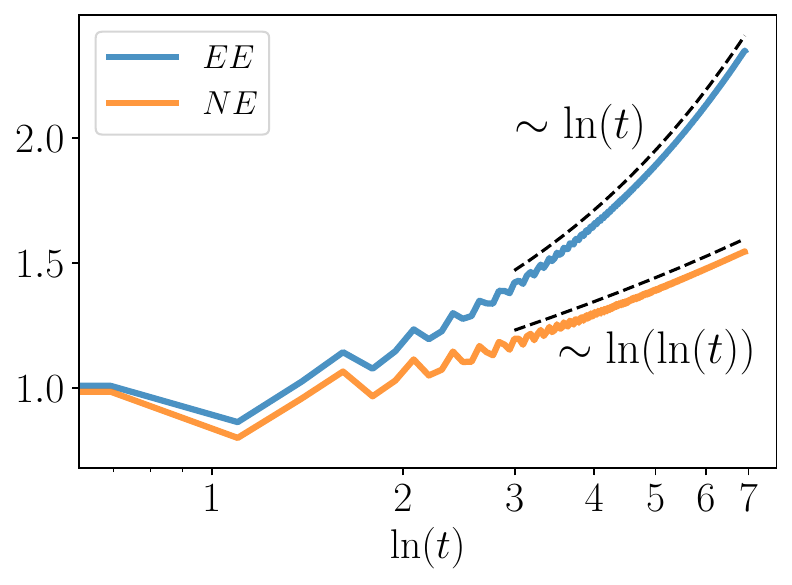}
    \caption{EE and NE as a function of $t$ for a homogeneous setup with no-defect whose Hamiltonian is given by Eq.~\eqref{supp_eq:Ham} with $g=g_c$. Note that the x-axis is $\ln(t)$ on a log scale. Since NE is a straight line, this means that it grows as $\ln(\ln(t))$. On the other hand, EE is exponential in this scale, which means that it grows as $\ln(t)$. Here, we have chosen $L_s = 320$ and $g = 0.5$. }
    \label{supp_fig:no_defect_Sn}
\end{figure}

In this section, we will present results for the evolution of the eigenvalue profile of the correlation matrix and the growth of NE in the absence of any defects and stochastic processes. We follow the same procedure as highlighted in the main text. We work with the free fermionic Hamiltonian, which is discussed in the main text, that is given by,
\begin{align}
\label{supp_eq:Ham}
    \hat{H} &= -g\sum_{\substack{i = -L_s\\ i\ne 0}}^{L} (\hat{c}^{\dagger}_i \hat{c}_{i+1} + \hat{c}^{\dagger}_{i+1} \hat{c}_{i}) - g_c (\hat{c}^{\dagger}_{0}\hat{c}_{1} + \hat{c}^{\dagger}_{1}\hat{c}_{0}) \notag\\
    &+A\sqrt{g^2 - g_c^2} \, (\hat{c}^{\dagger}_{0}\hat{c}_{0} - \hat{c}^{\dagger}_{1}\hat{c}_{1}).
\end{align}
In the absence of defect, i.e., $g = g_c$, we get the free fermionic Hamiltonian with homogeneous hopping $g$.
In Fig.~\ref{supp_fig:no_defect_eig_prof} we show the evolution of the system part of the correlation matrix. We begin with the domain wall initial condition. As a consequence of this, initially, all the eigenvalues are 1. As the system-reservoir setup is quenched, the particles flow from the system to the reservoir, which decreases the value of the eigenvalues of the system part of the correlation matrix. Contrary to the setup where we have a defect ($g\ne g_c$), which was discussed in the main text, here, most of the eigenvalues take values either 1 or 0. Hence, the leading order contribution to the EE or NE is 0. The contribution to the entropy comes from the $\mathcal{O}(1)$ number of the eigenvalues that are in transition from value 1 to value 0. From Fig.~\ref{supp_fig:no_defect_Sn} we observe that the NE in this case grows as $\ln(\ln(t))$, which is logarithmically slower than the EE, which scales as $\ln(t)$ with time \cite{CC2004,SKD2024}.

\section{Results for the hopping defect}
\label{sec:supp-HD}
\begin{figure}[h!]
    \centering
    \includegraphics[width=\linewidth]{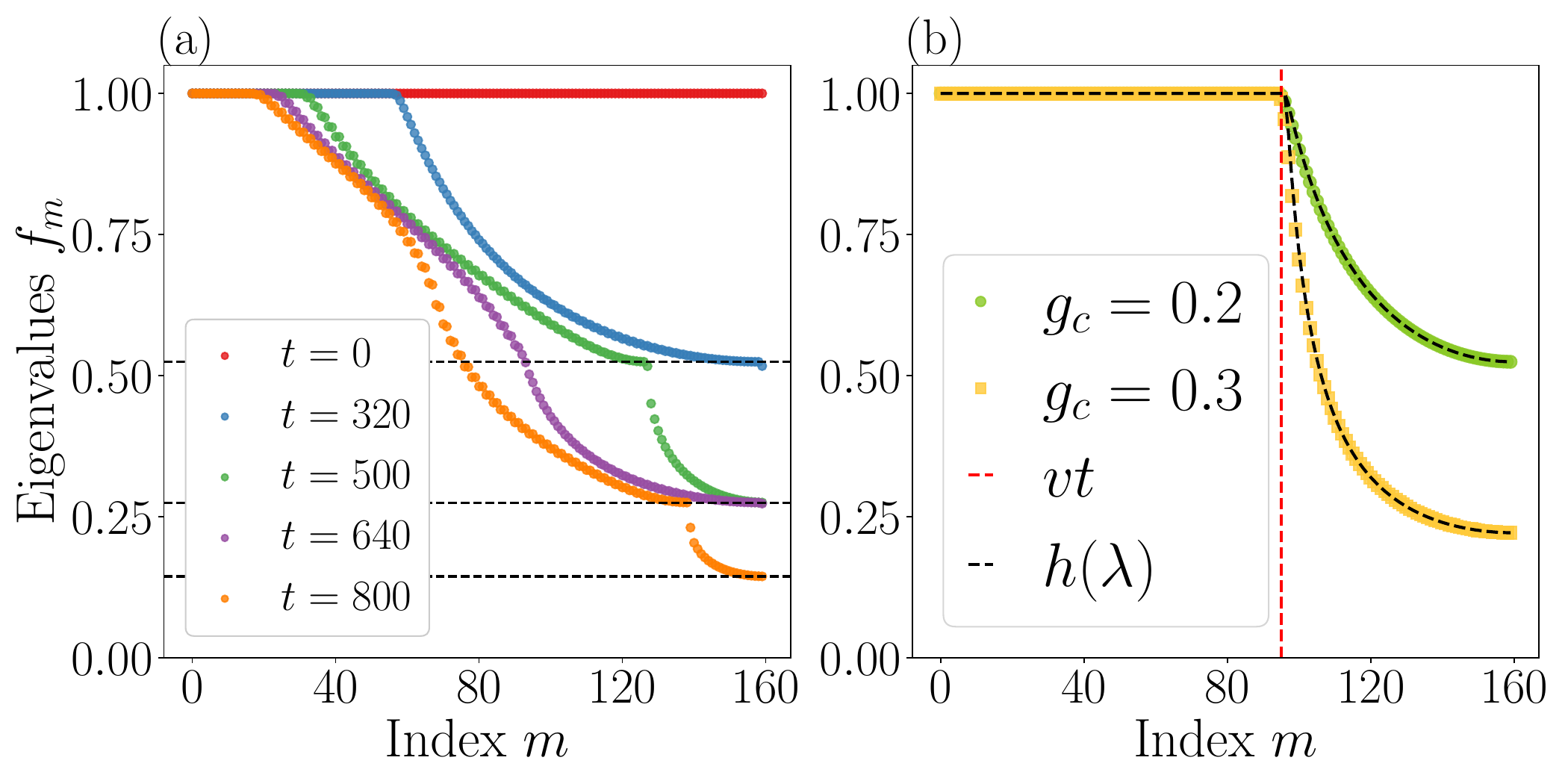}
    \caption{Eigenvalue profile of the system part of the correlation matrix in the case of a setup with hopping defect for $L_s = 160$ and $g = 0.5$ at (a) different times for $g_c= 0.2$. It can be seen that the eigenvalues show cascading (sharp drop) at $t = n\,L_s/g$ where $n = 1, 2, 3$.  The horizontal dashed black lines mark the minimum value of the sharp drop for each $n$. The value represented by the black line is given by, $\Big{[}\frac{(1-\lambda^2)^2}{(1+\lambda^2)^2}\Big{]}^n$.(b) Eigenvalue profile for different values of hopping constant of the defect $g_c$: $g_c = 0.2$ (green) and $g_c = 0.3$ (yellow) at time $t = 200$. The black dashed lines are represented by the function $h(\lambda)$, which is given by Eq.~\eqref{supp_eq:f_hopping} for the respective values of $\lambda = g_c/g$. The vertical red dashed line shows the position of the front for $t<L_s/g$, which is given by $L_s - vt$ where $v = 2g/\pi$.}  
    \label{supp_fig:eig_prof_hopp}
\end{figure}

\begin{figure}[h!]
    \centering
\includegraphics[width=0.9\linewidth]{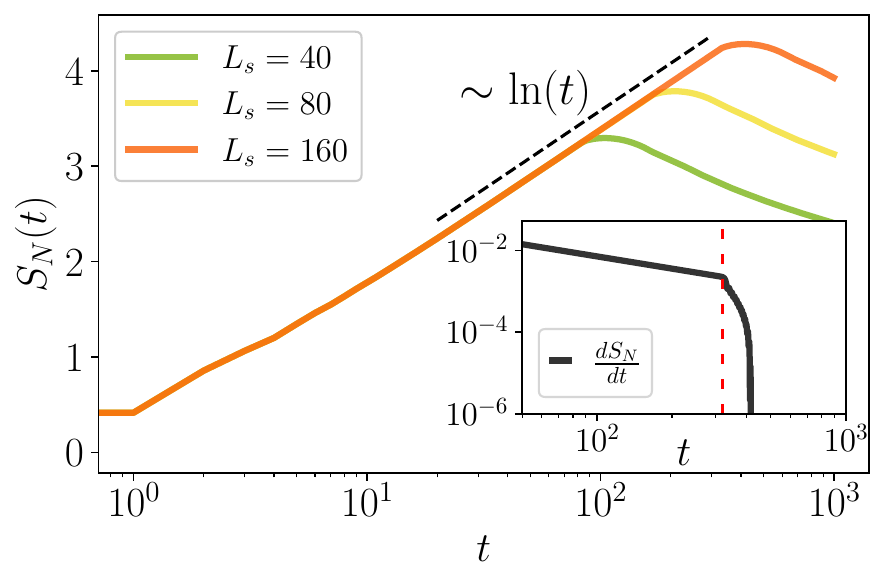}
    \caption{$S_N(t)$ vs $t$ for free fermionic setup with hopping defect which corresponds to $A = 0$ case in Eq.~\eqref{supp_eq:Ham}. Here, we have taken $g_c = 0.3$ and $g = 0.5$. NE shows a $\ln(t)$ growth with time. The inset shows $\frac{dS_N}{dt}$ as a function of time for $S_N(t)$ with $L_s = 160$, which is represented by the solid orange line in the main plot. The vertical dashed lines mark the time when non-analyticity is observed. The red vertical dashed line corresponds to $t = 320$.}
    \label{supp_fig:Hopp_SN}
\end{figure}

We recall here that in the main text, we discussed the case of free fermionic setup with a conformal defect. In this section, we will discuss the eigenvalue profile and the dynamics of NE for the setup with another type of defect, namely the hopping defect. The Hamiltonian in this case is given by Eq.~\eqref{supp_eq:Ham}
with $A = 0$. The eigenvalue profile for the hopping defect is shown in Fig.~\ref{supp_fig:eig_prof_hopp}. This profile shows similar qualitative behavior as in the case of the conformal defect (see Fig.~\ref{fig:eig_prof_conf}). Initially, at $t = 0$, all the eigenvalues are 1. Once the system-reservoir setup is quenched, an increasing number of eigenvalues take the value less than 1. Similar to the conformal defect case, a front moves from right to left with a velocity $v = 2g/\pi$, as seen in Fig.~\ref{supp_fig:eig_prof_hopp}b. Recall that the eigenvalue profile of the correlation matrix in the case of the conformal defect is observed to be closely related to the reflection probability, $R(k)$. But unlike the conformal defect case, $R(k)$ for hopping defect case depends on the momentum modes $k$. As a consequence, the eigenvalue distribution is expected to take a non-trivial form, which we confirm numerically. Upon inspection of the evolution of the eigenvalue profile, it can be observed that the eigenvalues of the system part of the correlation matrix, $f^{(h)}_m(t)$, take the form,
\begin{equation}
    f^{(h)}_m(t) = \begin{cases}
        & \hspace{3em} 1 \hspace{5em} \textrm{for}\,\, 1\le m\le L_s-vt\\
        &\frac{(1-\lambda^2)^2}{(1+\lambda^2)^2 - 4\lambda^2\Big{(}\frac{L_s-m}{vt}\Big{)}^2} \hspace{0.6em}\textrm{for}\,\, L_s-vt < m <L_s
    \end{cases}
    \label{supp_eq:f_hopping}
\end{equation}
where, $\lambda = g_c/g$ and as defined earlier, $v = 2g/\pi$. The superscript `$h$', in Eq.~\eqref{supp_eq:f_hopping}, denotes the hopping defect. As seen in Fig.~\ref{supp_fig:eig_prof_hopp}b, the eigenvalue profile described by Eq.~\eqref{supp_eq:f_hopping} is consistent with the numerical evolution. Note that the Eq.~\eqref{supp_eq:f_hopping} is valid only for $t<L_s/g$. Similar to the conformal defect case, we see a cascading of the eigenvalues at time $t = n\,L_s/g$, as shown in Fig.~\ref{supp_fig:eig_prof_hopp}a. Here $n$ is a natural number. Since the eigenvalue profile for the hopping defect case is not as trivial as the conformal defect case, analytically determining the $p(n,t)$ and the NE is not as straightforward. Instead, it can be determined numerically using the method prescribed in the main text. Fig.~\ref{supp_fig:Hopp_SN} shows the evolution of NE for the free fermionic system with the hopping defect for different system sizes. It is evident from the Fig.~\ref{supp_fig:Hopp_SN} that the NE increases as $\ln t$. Similar to the case of the conformal defect in the main text, when the finite size effect of the system kicks in, we observe non-analyticity in the evolution of NE. This non-analyticity is captured by the inset of Fig.~\ref{supp_fig:Hopp_SN}. The inset shows the evolution of rate of NE production as a function of time. We observe that it shows kink at the time that corresponds to the time when cascading of eigenvalues is observed in Fig.~\ref{supp_fig:eig_prof_hopp}a. We expect additional kinks at time $t = qL_s/g$ with $q>0$, which would be more pronounced in appropriate parameter regimes.

\section{Derivation of reflection and transmission coefficients}
\label{sec:supp-RT}

In the main text and section \ref{sec:supp-HD}, we have claimed that in the presence of a conformal defect or a hopping defect, the eigenvalue profile of the correlation matrix is closely related to the reflection probabilities of the plane wave eigenmodes that are scattered by the defect. In this section, for the sake of completeness, we will derive the relations for reflection [$R(k)$] and transmission [$T(k)$] probabilities for a plane wave propagating on a 1D free fermionic lattice that is scattered by a defect. We will derive the relation for $R(k)$ and $T(k)$ in the case of (i) conformal defect, and (ii) hopping defect. The free fermionic Hamiltonian with defect is given in Eq.~\eqref{supp_eq:Ham}
where  $A = 1$ corresponds to the conformal defect case and $A = 0$ corresponds to the hopping defect case. We want to work in a limit where the effect due to the finite size of the system or the reservoir does not come into play. Thus, we consider $L, L_s \xrightarrow{}\infty$. In this limit, the energy of the eigenmodes is given by $E(k) = -2g\cos(k)$. In a non-interacting free fermionic model, the defect acts like a barrier for the plane wave eigenmodes and thus reflects and transmits the plane wave with the momentum-dependent probabilities $R(k)$ and $T(k)$, respectively. The eigenstates in the presence of a defect take the form,
\begin{equation}
    \psi_k(x) = \begin{cases}
        & e^{ikx} + r(k) \, e^{-ikx} \hspace{3em} \textrm{for }x\le 0\\
        & t(k) \, e^{ikx} \hspace{6.6em} \textrm{for }x> 0,
    \end{cases}
    \label{supp_eq:eig_mode}
\end{equation}
where $r(k)$ and $t(k)$ are the coefficients to be determined, and they are related to reflection $[R(k)]$ and transmission $[T(k)]$ probabilities as $R(k)=|r(k)|^2$ and $T(k)=|t(k)|^2$. We will solve the eigenvalue equation, 
\begin{equation}
    h\,\psi_k = E(k)\,\psi_k
    \label{supp_eq:eigval_eq},
\end{equation} 
where $h$ is the single particle Hamiltonian of Eq.~\eqref{supp_eq:Ham} and $\psi_k$ are the new eigenmodes given in Eq.~\eqref{supp_eq:eig_mode}. This will give us a system of two linear equations, which upon solving, will give us $r(k)$ and $t(k)$.

\subsection{Conformal Defect}
In this subsection, we will solve the eigenvalue equation, given in Eq.~\eqref{supp_eq:eigval_eq}, for the conformal defect case, which corresponds to $A = 1$ in Eq.~\eqref{supp_eq:Ham}. We will use Eq.~\eqref{supp_eq:Ham} for $\hat{H}$ and Eq.~\eqref{supp_eq:eig_mode} for $\psi_k$ and solve for $x = 0$ and $x = 1$.
For $x = 0$, the eigenvalue equation [Eq.~\eqref{supp_eq:eigval_eq}] gives,
\begin{align}
        &-g\, \psi_k(-1) - g_c\, \psi_k(1) + \sqrt{g^2 - g_c^2}\,\,\psi_k(0) = E(k)\, \psi_k(0) \notag\\
    \label{eq:x=0}
\end{align}
and, for $x = 1$, the eigenvalue equation [Eq.~\eqref{supp_eq:eigval_eq}],  gives,
\begin{align}
    -g\, \psi_k(2) - g_c\, \psi_k(0) - \sqrt{g^2 - g_c^2}\,\,\psi_k(1) = E(k)\, \psi_k(1).
    \label{eq:x=1}
\end{align}
Substitute $\psi_k(x)$ from Eq.~\eqref{supp_eq:eig_mode}, using $E(k) = -2g\cos(k)$, and rearranging the equations, we get a system of linear equations for $r(k)$ and $t(k)$ as,
\begin{align}
\label{eq:LE1}
        &\hspace{-2em}-(e^{-ik} + \sqrt{1-\lambda^2})r(k) + \lambda e^{ik} \,t(k) = e^{ik} + \sqrt{1-\lambda^2},\\
        &\lambda\,r(k) - (1 - e^{ik}\sqrt{1-\lambda^2})\,t(k) = -\lambda,
        \label{eq:LE2}
\end{align}
where $\lambda = g_c/g$. The solution of Eq.~\eqref{eq:LE1} and Eq.~\eqref{eq:LE2} can be easily obtained as,
\begin{eqnarray}
    r(k) &=& -e^{ik}\sqrt{1-\lambda^2},\\
    t(k) &=& \lambda.
\end{eqnarray}
This gives the reflection and transmission probabilities as,
\begin{eqnarray}
\label{eq_supp:Rk_conf}
    R(k) &=& |r(k)|^2 = 1-\lambda^2,\\
    T(k) &=& |t(k)|^2 = \lambda^2.
\end{eqnarray}
It can be observed that in the case of a conformal defect, the reflection and transmission probabilities are independent of the momentum variable $k$. 

Comparing Eq.~\eqref{eq_supp:Rk_conf} with Eq.~\eqref{eq:f(k,t)} in the main text, we can clearly see that as the setup is quenched, upto time $t<L_s/g$, the value of the eigenvalues of the correlation matrix decreases and settles down to constant values that are equal to the $R(k)$. Furthermore, as the setup evolves in time, there is a cascading and saturation of the eigenvalues at the values given by $R(k)^{(q+1)}$ between time $t_q = qL_s/g$ and $t_{q+1} = (q+1)L_s/g$, where $q \in \mathbb{Z}$ and $q \ge 0$.
\\
\subsection{Hopping Defect}
In this subsection, we will obtain the equations for $R(k)$ and $T(k)$ for the case of the hopping defect, which corresponds to $A = 0$ in Eq.~\eqref{supp_eq:Ham}. We will follow the same procedure as in the conformal defect case, which involves solving the eigenvalue equation, Eq.~\eqref{supp_eq:eigval_eq}, by using Eq.~\eqref{supp_eq:Ham} for $\hat{H}$ and Eq.~\eqref{supp_eq:eig_mode} for $\psi_k$. The eigenvalue equation [Eq.~\eqref{supp_eq:eigval_eq}] for site $x = 0$ is given by,
\begin{align}
        &-g\, \psi_k(-1) - g_c\, \psi_k(1) = E(k)\, \psi_k(0)
    \label{supp_eq:x=0_hopping}
\end{align}
and, eigenvalue equation [Eq.~\eqref{supp_eq:eigval_eq}] for $x = 1$ is given by,
\begin{align}
    &-g\, \psi_k(2) - g_c\, \psi_k(0) = E(k)\, \psi_k(1).
    \label{supp_eq:x=0_hopping}
\end{align}
Again, substituting Eq.~\eqref{supp_eq:eig_mode} for $\psi_k$, using $E(k)=-2g\cos(k)$ and simplifying the expressions, we obtain a system of linear equations for $r(k)$ and $t(k)$ as,
\begin{eqnarray}
\label{eq:LE3}
    -e^{-ik}\,r(k) + \lambda\,e^{ik}\,t(k) &=& e^{ik}\\
    \lambda\,r(k) - t(k) &=& -\lambda,
    \label{eq:LE4}
\end{eqnarray}
where again we have used $\lambda = g_c/g$. Eq.~\eqref{eq:LE3} and Eq.~\eqref{eq:LE4} can be easily solved to get,
\begin{eqnarray}
    r(k) &=& \frac{1-\lambda^2}{\lambda^2 - e^{-2ik}},\\
    t(k) &=& \frac{2i\lambda\sin(k)}{\lambda^2 e^{ik} - e^{-ik}}.
\end{eqnarray}
This gives reflection and transmission probabilities as,
\begin{eqnarray}
\label{eq_supp:Rk_hopp}
    R(k) &=& |r(k)|^2 = \frac{(1-\lambda^2)^2}{(1+\lambda^2)^2 - 4\lambda^2\cos(k)},\\
    T(k) &=& |t(k)|^2 = \frac{4\lambda^2\sin^2(k)}{(1+\lambda^2)^2 - 4\lambda^2\cos(k)}.
\end{eqnarray}
We compare $R(k)$ in Eq.~\eqref{eq_supp:Rk_hopp} with $f^{(h)}_m(t)$ in Eq.~\eqref{supp_eq:f_hopping}. Upon quenching, for $t<L_s/g$, the value of the eigenvalues of the correlation matrix takes the form of a function that is related to the $R(k)$. We identify the $\cos(k)$ in Eq.~\eqref{eq_supp:Rk_hopp} with $\Big{(}\frac{L_S-m}{vt}\Big{)}$ in Eq.~\eqref{supp_eq:f_hopping}.

\end{document}